\documentclass{article}
\usepackage{indentfirst}
\usepackage{bm}
\usepackage{epsfig}
\usepackage{epstopdf, graphics, graphicx}
\usepackage{bm, amsmath}
\usepackage{multirow}
\usepackage{longtable}
\usepackage{balance}
\usepackage{lastpage}
\usepackage[utf8]{inputenc}
\usepackage[colorlinks=true, urlcolor=blue, pdffitwindow=true, linkcolor=blue, citecolor=blue, pdfstartview={FitH}]{hyperref}
\usepackage[super]{cite}

\begin{document}

\begin{center}
\large\bf Iterative density matrix revisited:\\ Excitonic phenomena to InAs quantum well as a saturable absorber 
\end{center}

%\footnotetext{\hspace*{-.45cm}\footnotesize $^*...........$.}
\footnotetext{\hspace*{-.45cm}\footnotesize $^{*}$\sl E-mail: sami.ortakaya@yahoo.com}

\begin{center}
\rm Sami Ortakaya$^{\rm{a}, \rm{b}, *}$
\end{center}

\begin{center}
\begin{footnotesize} \sl
${}^{\rm a)}$\it Ercis Central Post Office, 65400 Van, Turkey\\
${}^{\rm b)}$Shipito address: 444 Alaska Avenue Suite $\#$BKF475
Torrance, CA 90503 USA\\
\end{footnotesize}
\end{center}

\vspace*{2mm}

\begin{center}
\begin{minipage}{13cm}
\textbf{Abstract.}By solving the Liouville equation, third-order nonlinear terms is found via iterative density matrix. Regarding the improved modeling, all frequency range is taken instead of weak absorptive limit. Considered process can be compared with saturation fitting for heavy-hole excitons in the InAs quantum well.
\end{minipage}
\end{center}

\begin{center}
\begin{minipage}{13cm}{\bf Keywords:}
exciton, dielectric constant, InAs quantum well, density matrix approach, third-order nonlinear phenomena 
\end{minipage}
\end{center}

\section{Introduction}
In addition to the fact that optical coefficients depend on the particle population, the required optical gain plays a key role in electroabsorptive quantum devices. The electrical polarization scheme, which is related to the electrical field of the incident optical wave, leads to the dielectric permittivity in the quantum confinement space. It is well known that the dielectric constant is obtained from the dipole moment per unit volume for the dipole oscillator \cite{foxvb}. Not only does a classical oscillator with a forced system yield polarization, but optical transitions during the time evolution of quantum states also contribute to polarization. In this context, the mean dipole moment can be calculated from the time evolution of the density matrix. An iterative method is used for nonlinearities in the density matrix elements \cite{chuang}. In a way, the third-order susceptibility resulting from this iterative method leads to third-order absorption, which has negative values. Nevertheless, we can observe decreasing absorption with increasing optical intensity due to these negative values. A familiar approach to modeling intensity-dependent absorption is through a saturable absorber, which is obtained from \cite{fox2}
\begin{equation}
\alpha(I)=\frac{\alpha_{0}}{1+I/I_{s}}
\label{sat}\end{equation}
where $\alpha_{0}$ denotes linear term, $I_{s}$ and $I$ are saturable and variable intensities, respectively. 

A lot of research has been conducted on the nonlinear term in the context of quantum confinement and size effects \cite{s1, s2, s4, s5, s6, s7}. Since the negative values can affect the total optical coefficients, many studies have focused on the dielectric function within the Lorentzian limit. Heterocrystal size effects have been used to model this limitation. Both the linear and nonlinear terms lead to Lorentzian peaks in absorption coefficients when weak absorptive considerations are taken into account. There are several relations between excitonic phenomena and the electron-to-electron regime \cite{a1, a2, a3, a4}. Additionally, the dielectric function with Lorentzian broadening has been calculated in the context of quantum confinement \cite{diel1}. While the iterative approach to the density matrix leads to negative values and decreasing absorption coefficients, we can still deal with the entire frequency range without any weak absorber limit. Through iterative considerations, polarization can be expanded to high orders, and it can be observed that the optical coefficients depend on the optical intensity which is given in the form
\begin{equation}
\alpha(I)=\sqrt{a+I\cdot b}.
\label{sat3}\end{equation}
Here, $a$ and $b$ are independent constants that assume the nonlinearity of the "equal frequency output" is valid for $\omega + 0\to\omega$.

\section{Solutions of the Liouville equation for dipole interaction}
In order to obtain mean dipole moment for considered optical coefficients, we take that the optical wave has electric field $\mathcal{E}_{0}$ which interact with single particles. Interaction Hamiltonian as a perturbed term, can be given in the form
\begin{eqnarray}
H_{\rm int}=-M\mathcal{E}_{0}=\left(\begin{array}{ccccc} 
M_{11}\mathcal{E}_{0} & M_{12}\mathcal{E}_{0} & \cdots & M_{1 f}\mathcal{E}_{0} & \cdots\\ 
M_{21}\mathcal{E}_{0} & M_{22}\mathcal{E}_{0} & \cdots & M_{2 f}\mathcal{E}_{0} & \cdots\\
\cdots\ & \cdots\ & \cdots\ & \cdots\ & \cdots\\
M_{ f1}\mathcal{E}_{0} & M_{ f2}\mathcal{E}_{0} & \cdots & M_{ ff}\mathcal{E}_{0} & \cdots\\
\cdots\ & \cdots\ & \cdots\ & \cdots\ & \cdots
 \end{array}\right),
\end{eqnarray}
 where dipole matrix element is defined by $M_{if}=\left<\Psi_{i}|er|\Psi_{f}\right>$ which pure quantum states $\Psi_{i}$ and $\Psi_{f}$. Mean value of the dipole moment in unit volume $V$ is useful for the polarization, so a monochromatic wave of frequency $\omega$ yields polarization of the form $P(\omega)=\frac{\left<M\right>}{V}.$ Considering density matrix related to the mean dipole moment, we should have
\begin{equation}
\left<M\right>=\sum_{i,\,f}\left<f|\rho|i\right> \left<i|M|f\right>={\rm Tr}(\rho M),
\end{equation}
where density matrix is defined by
\begin{eqnarray}
\rho=\left(\begin{array}{ccccc} 
\rho_{11} & \rho_{12} & \cdots & \rho_{1 f} & \cdots\\ 
\rho_{21} & \rho_{22} & \cdots & \rho_{2 f} & \cdots\\
\cdots\ & \cdots\ & \cdots\ & \cdots\ & \cdots\\
\rho_{f1} & \rho_{f2} & \cdots & \rho_{f f} & \cdots\\
\cdots\ & \cdots\ & \cdots\ & \cdots\ & \cdots
 \end{array}\right).
\end{eqnarray}
Quantum levels including unperturbed Hamiltonian $H_0$ within time-$t$ evolution behave as the following Liouville's equation \cite{Shen2003}
\begin{equation}
\frac{\partial\rho(\omega,\,t)}{\partial t}=-\frac{{\rm i}}{\hbar}\left[H_{0} - M\mathcal{E}_{0},\,\rho\right]-\Gamma\rho-\rho\Gamma,
\label{liouville}
\end{equation}
where $\Gamma$ denotes ''damping'' related to scattering and phonon interactions. Within framework of the matrix-trace operation, we should write
{\setlength\arraycolsep{0.1em}\begin{eqnarray}
P(\omega,\,t)=\frac{1}{V}{\rm Tr}(\rho M)=\frac{1}{V}\big[\rho_{11}M_{11}+\rho_{12}M_{21}+
\rho_{21}M_{12}+\rho_{22}M_{22}\big].
\label{trace}\end{eqnarray}}
From Equation (\ref{liouville}), we obtain that $\rho_ {21}$ has the following form
{\setlength\arraycolsep{0.1em}\begin{eqnarray}
\frac{\partial \rho_{21}}{\partial t}&=&\Big<2\Big|\frac{\partial \rho}{\partial t}\Big|1\Big>\nonumber\\
&=&\frac{-{\rm i}}{\hbar}
\Big[\left<2|(H_{0}-M\mathcal{E}_0)\rho|1\right>-\left<2|\rho(H_{0}-M\mathcal{E}_0)|1\right>\Big]-\Big[\left<2|\Gamma\rho|1\right>+\left<2|\rho\Gamma|1\right>\Big]\nonumber\\
&=&\frac{-{\rm i}}{\hbar}
\Big[\big<2\big|H_{0}(|1\big>\big<1|+|2\big>\big<2|)\rho\big|1\big>-\big<2\big|M\mathcal{E}_{0}(|1\big>\big<1|+|2\big>\big<2|)\rho\big|1\big>\Big]
\nonumber\\
&+&\frac{{\rm i}}{\hbar}
\Big[\big<2\big|\rho(|1\big>\big<1|+|2\big>\big<2|)H_{0}\big|1\big>-\big<2\big|\rho(|1\big>\big<1|+|2\big>\big<2|)M\mathcal{E}_{0}\big|1\big>\Big]\nonumber\\
&-&\Big[\left<2|\Gamma(|1\big>\big<1|+|2\big>\big<2|)\rho|1\right>+\left<2|\rho(|1\big>\big<1|+|2\big>\big<2|)\Gamma|1\right>\Big],
\end{eqnarray}}
where eigenvalues $E_1$ ve $E_2$ provide that $\left<1|H_0|1\right>$=$E_ {1}$, $\left<2|H_0|2\right>$=$E_ {2}$. Using that non-diagonal elements are $\left<2|H_0|1\right>$=0, $\left<1|H_0|2\right>$=0, we have
{\setlength\arraycolsep{0.1em}\begin{eqnarray}
\frac{\partial \rho_{21}}{\partial t}=\frac{-{\rm i}}{\hbar}
\Big[(E_{21}+(\Delta M)\mathcal{E}_{0})\rho_{21}+(\rho_{22}-\rho_{11})M_{21}\mathcal{E}_{0}\Big]-(\Gamma_{22}+\Gamma_{11})\rho_{21}-\Gamma_{21}(\rho_{11}+\rho_{22}).\nonumber\\
\label{ilke}\end{eqnarray}}
Here,  $E_{21}=E_2 - E_1$ and $\Delta M=M_{11}-M_{22}$. The formation regarding phenomenological operator for damping is defined relaxation time via non-diagonal and diagonal elements for $\Gamma_{21}$=$\Gamma_{12}$=0 and $\Gamma_{11}=\Gamma_{22}=\gamma/2$, respectively. So that, Equation (\ref{ilke}) reads
{\setlength\arraycolsep{0.1em}\begin{eqnarray}
\frac{\partial \rho_{21}}{\partial t}=\frac{-{\rm i}}{\hbar}
\Big[(E_{21}+(\Delta M)\mathcal{E}_{0}-{\rm i}\hbar\gamma)\rho_{21}+(\rho_{22}-\rho_{11})M_{21}\mathcal{E}_{0}\Big].
\end{eqnarray}}
In the presence of the monochromatic optical wave of electric field $\mathcal{E}=\mathcal{E}_{0}\cos\omega t$, we consider the same output frequency, then we get
\begin{equation}P_{\rm res}(\omega,\, t)=\left(\chi^{(1)}\mathcal{E}_{0}+
\frac{3}{4}\chi^{(3)}\mathcal{E}_{0}^{3}+\cdots\right){\rm e}^{{\rm- i}\omega t},\end{equation}
where the key role is to get the equal output frequency as $\omega+0\to\omega$, so we take that the stationary solution becomes
\begin{equation}
\rho(\omega,\,t)=\rho(\omega){\rm e}^{{\rm- i}\omega t}.\label{cozgen1}
\end{equation}
Inserting the proposed ways, matrix elements lead to
\begin{subequations}
{\setlength\arraycolsep{0.1em}\begin{eqnarray}
\rho_{21}({\omega})&=&\frac{(\rho_{11}-\rho_{22})M_{21}\mathcal{E}_{0}}{E_{21}-\hbar\omega+(\Delta M)\mathcal{E}_{0}-{\rm i}\hbar\gamma},\\
\rho_{12}({\omega})&=&\frac{(\rho_{11}-\rho_{22})M_{12}\mathcal{E}_{0}}{E_{21}+\hbar\omega+(\Delta M)\mathcal{E}_{0}+{\rm i}\hbar\gamma},\\
\rho_{11}(\omega)&=&
\frac{(M_{21}\rho_{12}-M_{12}\rho_{21})\mathcal{E}_{0}}{\hbar\omega+{\rm i}\hbar\gamma},\\
\rho_{22}(\omega)&=&
\frac{(M_{12}\rho_{21}-M_{21}\rho_{12})\mathcal{E}_{0}}{\hbar\omega+{\rm i}\hbar\gamma}.
\end{eqnarray}}
\end{subequations}
From Equation (\ref{trace}), polarization of the optical transition leads to
{\setlength\arraycolsep{0.1em}\begin{eqnarray}
P_{\rm res}(\omega)=\frac{(\rho_{11}-\rho_{22})|M_{12}|^{2}}{V}\bigg[\mathcal{E}_{0}\left(\frac{1}{E_{+}}+\frac{1}{E_{-}}\right)+(\Delta M)\mathcal{E}_{0}^{2}\left(\frac{1}{\hbar\omega+{\rm i}\hbar\gamma}\right)\left(\frac{1}{E_{+}}-\frac{1}{E_{-}}\right)\bigg],\nonumber\\
\label{anadenklem1}
\end{eqnarray}} 
where, $E_{-}=E_{21}-\hbar\omega+(\Delta M)\mathcal{E}_{0}-{\rm i}\hbar\gamma$ and $E_{+}=E_{21}+\hbar\omega+(\Delta M)\mathcal{E}_{0}+{\rm i}\hbar\gamma$. Optical radiation of frequency $\omega$ cause to the resonance polarization
\begin{equation}
P_{\rm res}(\omega,\,t)=\epsilon_{0}\Big[A_{1}\chi_{\rm res}^{(1)}(\omega)\mathcal{E}_{0}+A_{3}\chi_{\rm res}^{(3)}(\omega)\mathcal{E}_{0}^ 3+A_{5}\chi_{\rm res}^{(5)}(\omega)\mathcal{E}_{0}^5 +\cdots\Big]{\rm e}^{{\rm- i}\omega t},
\end{equation}
where $A_{1}$=1, $A_{3}$=$\frac{3}{4}$ and $A_{5}$=$\frac{5}{8}$ are determined from trigonometric ways.

{\textit{ Linear term.}} We consider firstly that Equation (\ref{anadenklem1}) includes
$$P_{\rm res}(\omega,\,t)\to\epsilon_{0}\chi_{\rm res}^{(1)}(\omega)\mathcal{E}_{0}{\rm e}^{{\rm- i}\omega t}$$
and occupancy of the subbands can be proposed as
$$\rho_{11}-\rho_{22}\to\rho_{11}^{(0)}-\rho_{22}^{(0)}=(N_{\rm el}+1)-(N_{\rm el})=1$$
where $N_{\rm el}$ is particle number. Typically, we take that $N_{\rm el}$=1 denotes full-occupied and empty-excited states. Then, we should get the result in the form
\begin{equation}
\chi_{\rm res}^{(1)}(\omega)=\frac{|M_{12}|^2}{\epsilon_{0}V}\left[\frac{1}{E_{+}}+\frac{1}{E_{-}}+\frac{(\Delta M)\mathcal{E}_{0}}{\hbar\omega+{\rm i}\hbar\gamma}\left(\frac{1}{E_{+}}-\frac{1}{E_{-}}\right)\right].
\end{equation}
{\textit{ Nonlinearities.}} Equation (\ref{anadenklem1}) is written by
$$P_{\rm res}(\omega,\,t)\to\frac{3\epsilon_0}{4}\chi_{\rm res}^{(3)}(\omega)\mathcal{E}_{0}^{3}{\rm e}^{{\rm- i}\omega t}$$
As a second iteration, diagonal elements yield
{\setlength\arraycolsep{0.1em}\begin{eqnarray}
\rho_{11}-\rho_{22}\to\rho_{11}^{(1)}-\rho_{22}^{(1)}&=&\frac{2(M_{21}\rho_{12}^{(1)}-M_{12}\rho_{21}^{(1)})\mathcal{E}_{0}}{\hbar\omega+{\rm i}\hbar\gamma}\nonumber\\
&=&2|M_{12}|^2 \mathcal{E}_{0}^{2}\left(\frac{(\rho_{11}^{(0)}-\rho_{22}^{(0)})}{\hbar\omega+{\rm i}\hbar\gamma}\right)\left(\frac{1}{E_{+}}-\frac{1}{E_{-}}\right)\nonumber\\
&=&\left(\frac{2|M_{12}|^2 \mathcal{E}_{0}^{2}}{\hbar\omega+{\rm i}\hbar\gamma}\right)\left(\frac{1}{E_{+}}-\frac{1}{E_{-}}\right),
\end{eqnarray}} 
so we obtain that third order term as follows:
{\setlength\arraycolsep{0.1em}\begin{eqnarray}
\chi_{\rm res}^{(3)}(\omega)=\frac{8|M_{12}|^4}{3\epsilon_{0}V}\frac{1}{\hbar\omega+{\rm i}\hbar\gamma}\left[\left(\frac{1}{E_{+}^2}-\frac{1}{E_{-}^2}\right)+\frac{(\Delta M)\mathcal{E}_{0}}{\hbar\omega+{\rm i}\hbar\gamma}\left(\frac{1}{E_{+}}-\frac{1}{E_{-}}\right)^2\right].
\end{eqnarray}} 
For $j=2N+1$ and $(N=0,\,1,\,2,\,3,\dots)$ values, general form of the polarization and susceptibility with $j$'th order becomes
{\setlength\arraycolsep{0.1em}\begin{eqnarray}P^{(j)}_{\rm res}(\omega)&=&A_{j}\epsilon_{0}\chi^{(j)}_{\rm res}(\omega)\mathcal{E}_{0}^{j}\nonumber\\
&=&\frac{\rho_{11}^{(\frac{j-1}{2})}-\rho_{22}^{(\frac{j-1}{2})}}{V}
|M_{12}|^{2} \mathcal{E}_{0}\left[
\frac{1}{E_{-}}+\frac{1}{E_{+}}+\frac{(\Delta M)\mathcal{E}_0}{\hbar\omega+{\rm i}\hbar\gamma} \left(\frac{1}{E_{+}}-\frac{1}{E_{-}}\right)
\right]\nonumber\\
&=&\left(\frac{2^{\frac{j-1}{2}} |M_{12}|^{j+1} \mathcal{E}_{0}^{j}}{V}\right)\left(
\frac{\frac{1}{E_{+}}-\frac{1}{E_{-}}}{\hbar\omega+{\rm i}\hbar\gamma}\right)^{\frac{j-1}{2}}\left[
\frac{1}{E_{-}}+\frac{1}{E_{+}}+\frac{(\Delta M)\mathcal{E}_0}{\hbar\omega+{\rm i}\hbar\gamma} \left(\frac{1}{E_{+}}-\frac{1}{E_{-}}\right)
\right],\nonumber\\
&&\\
\chi^{(j)}_{\rm res}(\omega)&=&\frac{2^{\frac{j-1}{2}} |M_{12}|^{j+1} }{A_j \epsilon_{0}V}\left(
\frac{\frac{1}{E_{+}}-\frac{1}{E_{-}}}{\hbar\omega+{\rm i}\hbar\gamma}\right)^{\frac{j-1}{2}}\left[
\frac{1}{E_{-}}+\frac{1}{E_{+}}+\frac{(\Delta M)\mathcal{E}_0}{\hbar\omega+{\rm i}\hbar\gamma} \left(\frac{1}{E_{+}}-\frac{1}{E_{-}}\right)
\right],
\end{eqnarray}}
where iterative difference is obtained as
{\setlength\arraycolsep{0.1em}\begin{eqnarray}
\rho_{11}-\rho_{22}\to\rho_{11}^{(\frac{j-1}{2})}-\rho_{22}^{(\frac{j-1}{2})}&=&2^{\frac{j-1}{2}} |M_{12}|^{j-1} \mathcal{E}_{0}^{j-1}\left(
\frac{\frac{1}{E_{+}}-\frac{1}{E_{-}}}{\hbar\omega+{\rm i}\hbar\gamma}\right)^\frac{j-1}{2}.\end{eqnarray}}
\section{Relative Dielectric Permittivity}
We would like to get the transition between allowed states instead of the classical dipole resonance, so we re-write resonance context ("res") in the dielectric constant which depends on the susceptibilities as follows:
\begin{equation}
\tilde{\epsilon}_{r}=1+\chi^{(1)}+\tilde{\chi}^{(1)}_{\rm res}.
\end{equation}
For linear term, dielectric constant becomes the formation which is given by
{\setlength\arraycolsep{0.1em}\begin{eqnarray}
\tilde{\epsilon}_{r}&=&1+\chi^{(1)}+\tilde{\chi}^{(1)}_{\rm res}\nonumber\\
&=&1+\chi^{(1)}+\frac{|M_{12}|^2}{\epsilon_{0}V}\left[\frac{1}{E_{+}}+\frac{1}{E_{-}}+\frac{(\Delta M)\mathcal{E}_{0}}{\hbar\omega+{\rm i}\hbar\gamma}\left(\frac{1}{E_{+}}-\frac{1}{E_{-}}\right)\right].\label{dielr}
\end{eqnarray}}
Note that the key features which cause to the spectra at high frequencies (it is limited to infrared-visible range) is to constitute overlap of allowed states and energy gap.
\subsection{Third-order nonlinearity}
Dielectric displacement regarding absorptive material is given by
\begin{equation}
\mathbf{D}=\epsilon_{0}\mathcal{E}+\mathbf{P}=\epsilon_{0}\mathcal{E}+\epsilon_{0}\chi^{(1)}\mathcal{E}+\epsilon_{0}\tilde{\chi}^{(1)}(\omega)\mathcal{E}+\frac{3}{4}\epsilon_{0}\tilde{\chi}^{(3)}(\omega)\mathcal{E}^3,
\end{equation}
where we deal with the output, $\omega+0\to\omega$. Due to the displacement
$\mathbf{D}=\epsilon_{0}\epsilon_{r}\mathcal{E}$, nonlinear dielectric constant becomes
\begin{equation}\epsilon_{r}^{\rm NL}=\epsilon_{\infty}+\tilde{\chi}^{(1)}(\omega)+\frac{3}{4}\tilde{\chi}^{(3)}(\omega)\mathcal{E}_{0}^2,\qquad\epsilon_{\infty}=1+\chi^{(1)}.\label{nelep}
\end{equation}
We consider that the optical intensity relation which given as $\mathcal{E}_{0}^2=\frac{2I}{n_{0}\epsilon_{0}c}$, then we have
\begin{equation}
\epsilon_{r}^{\rm NL}(I)=a+I\cdot b,
\label{sat2}\end{equation}
where $a$ and $b$ are intensity-independent constants, $n_{0}$, $\epsilon_{0}$ and $c$ are refractive index, dielectric permittivity and speed of light under free space, respectively. Inserting dielectric constant to the complex refractive index, real and imaginary parts are given by \cite{fox2}
\begin{subequations}
{\setlength\arraycolsep{0.1em}\begin{eqnarray}{\rm Re}[\tilde{n}]&=&\sqrt{\frac{\epsilon_{1}^2+
\left(\epsilon_{1}^2+\epsilon_{2}^2\right)^{1/2}}{2}},\\
{\rm Im}[\tilde{n}]&=&\sqrt{\frac{-\epsilon_{1}^2+
\left(\epsilon_{1}^2+\epsilon_{2}^2\right)^{1/2}}{2}},
\end{eqnarray}}
\end{subequations}
where $\epsilon_{1}$=${\rm Re}[\tilde{\epsilon}_{r}]$ and $\epsilon_{2}$=${\rm Im}[\tilde{\epsilon}_{r}]$ represent real and imaginary parts, respectively. For the absorption coefficient spectra, we get 
\begin{equation}
\alpha(\omega)=\frac{2\omega{\rm Im}[\tilde{n}]}{c}.
\end{equation}
Intensity-dependent dielectric constant in Equation (\ref{nelep}) might provide decreasing with increasing intensity which is given in Eqs. (\ref{sat}) and (\ref{sat3}).
\section{Numerical Results for Unstrained InAs Quantum Well}
We examine InAs quantum well (QW) under intraband transitions via intensity dependent coefficients. Firstly, there is well known 2D-polar excitonic space under Coulomb interaction through infinite well. By defining variational parameter $\xi$ and QW-width $z$, wavefuction and binding energy of the exciton are given by \cite{a1}
\begin{equation}
\varphi_{\rm{ex}}({\bm r})=\sqrt{\frac{2}{\pi}}\frac{1}{\xi}\exp\left(-\frac{r}{\xi}\right),\quad E_{\rm b}=-\frac{\hbar^2}{2\mu\xi^2}+\frac{e^2}{4\pi\epsilon_{st}}\left<\Psi_{\rm ex}\left|\frac{1}{\sqrt{(z_{\rm e}-z_{\rm h})^2+r^2}}\right|\Psi_{\rm ex}\right>.
\end{equation}
Here, $\xi$ is also known as exciton Bohr radius, $\mu$ denotes reduced mass of electron and hole. The general formation satisfies that $$\Psi_{\rm ex}(z_{\rm e},\,z_{\rm h},\,r)=\varphi_{\rm e}\varphi_{\rm h}\varphi_{\rm ex}(r)$$
with quantum confinement equations $H_{\rm e}\varphi_{\rm e}$=$E_{\rm e}\varphi_{\rm e}$ and $H_{\rm h}\varphi_{\rm h}$=$E_{\rm h}\varphi_{\rm h}$ for electron and hole, respectively. Typical quantum confinements permit our discussion in more simply lines, so we can also use $\bm k\cdot\bm p$ perturbation via formation
\begin{equation}
|\left<c|\bm p|v\right>|^2\cdot|\left<\varphi_{\rm_h}|\varphi_{\rm e}\right>|^2=\frac{m_{0}^2 E_{g}}{2m^{*}_{e}}\cdot|\left<\varphi_{\rm_h}|\varphi_{\rm e}\right>|^2,
\end{equation}
we also conclude that the dipole approximation on the momentum matrix yields 
\begin{equation}|M_{12}|^2=|\left<1|e\cdot r|2\right>|^2=\frac{\hbar^2}{E_{\rm ex}^2m_{0}^2}|\left<c|\bm p|v\right>|^2\cdot|\left<\varphi_{\rm_h}|\varphi_{\rm e}\right>|^2,\end{equation}
where $E_{\rm ex}$=$E_{\rm e}+E_{\rm h}-E_{\rm b}$. 

\begin{figure*}[!hbt]\centering
\scalebox{1}{\includegraphics{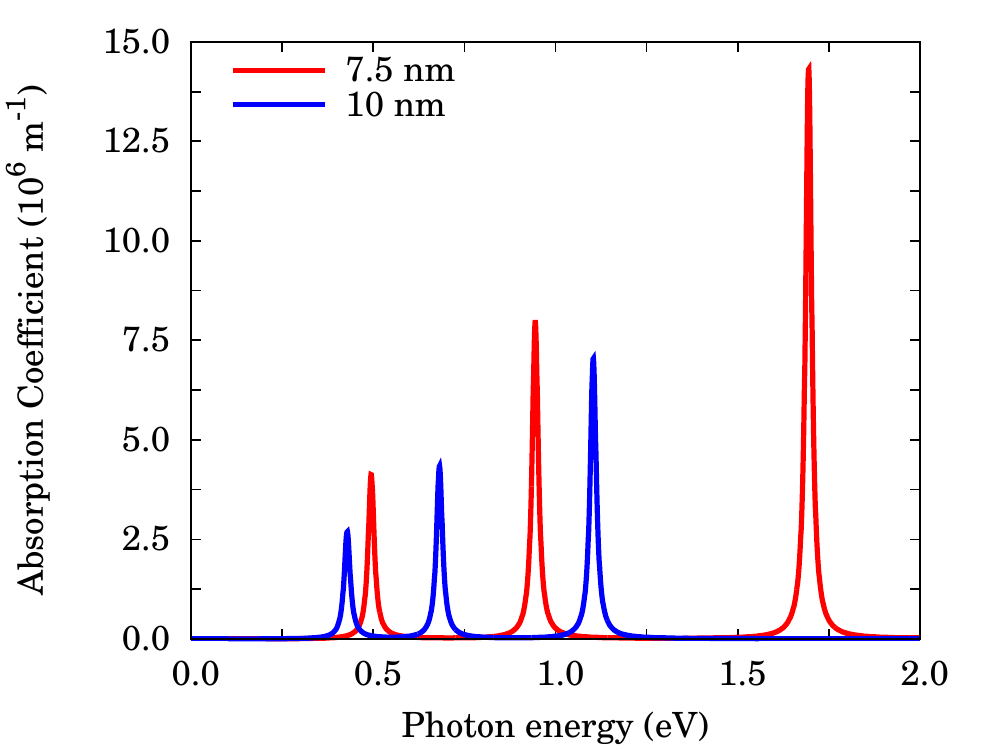}}
\scalebox{1}{\includegraphics{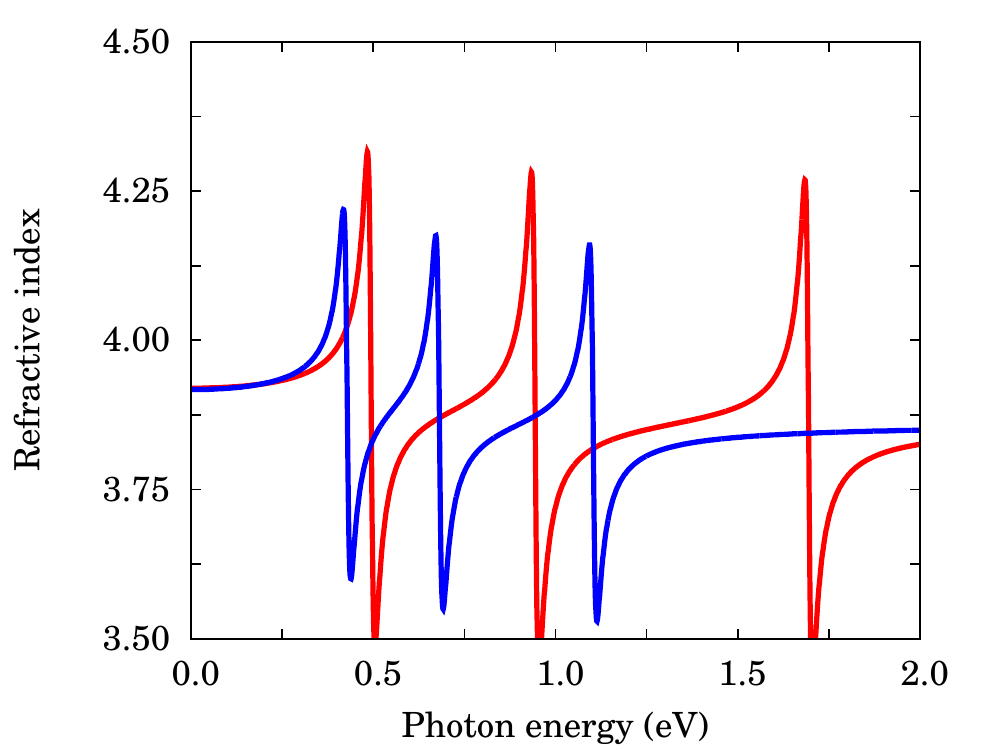}}
\caption{Liner optical coefficients related to the hh excitonic transitions on the typical InAs QW. Here, we take that $E_{g}$=0.325 eV.}\label{abs}
\end{figure*}
\begin{figure}[!hbt]
\centering
\setbox1=\hbox{\includegraphics[height=5cm]{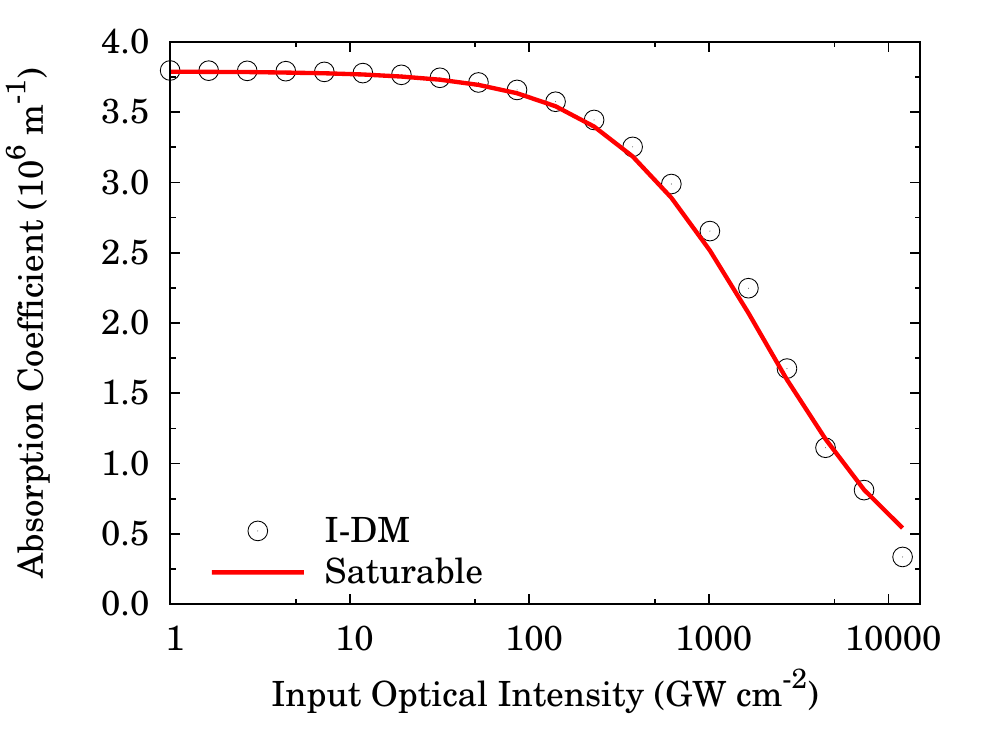}}
\includegraphics[height=8cm]{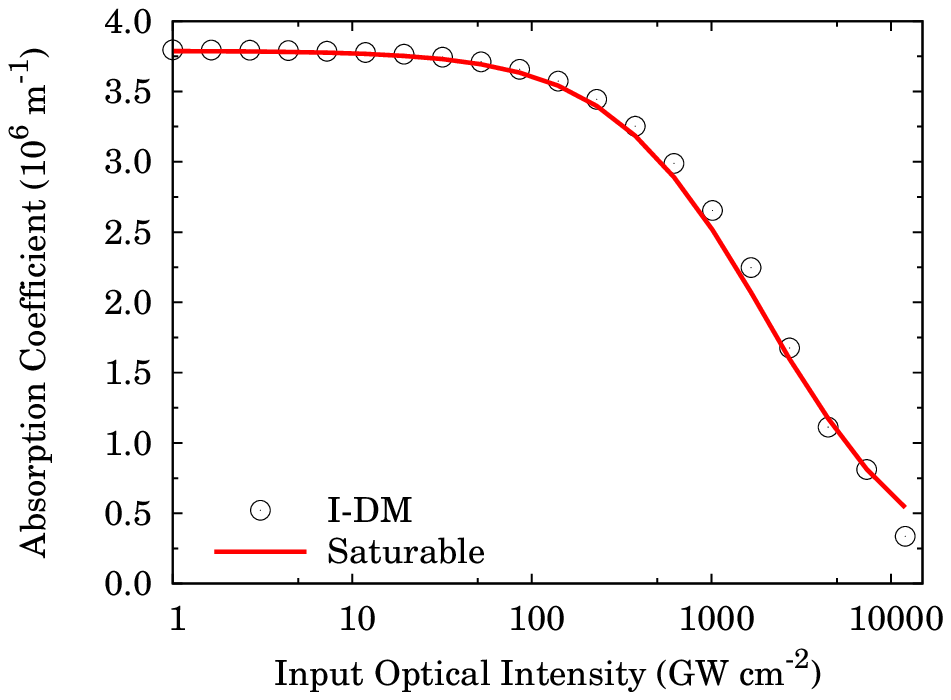}\llap{\makebox[\wd1][l]{\raisebox{2.8cm}{\includegraphics[height=2.8cm]{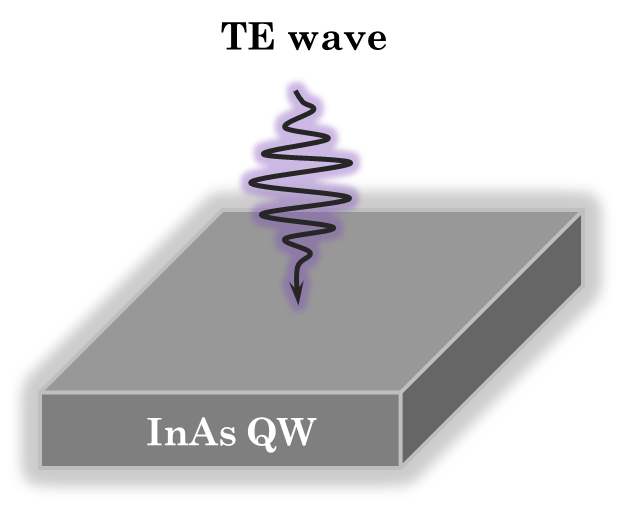}}}}
\caption{Comparative decreasing for the optical saturation fitting and iterative density matrix (I-DM) which are given in Equations (\ref{sat}) and (\ref{sat3}), respectively. }\label{sat4}
\end{figure}

Figure \ref{abs} shows that intraband optical transitions occur at infrared frequencies. The intrinsic Fermi level is taken at 300 K, and the occupancy depends on the heavy-hole (hh) exciton, so we obtain $1/V=2(f_{1}-f_{2})/L_{w}$, where $L_w$ is the quantum well width and $f_1$ and $f_2$ are the Fermi-Dirac (FD) distribution functions. The density of states takes the form $\mu/\pi\hbar^2$ for the hh exciton. All transitions occur with TE-polarized optical waves, and we neglect light-hole exciton transitions due to their small values (1/3 factor of TE modes). The static dielectric constant is calculated to be 15.058 and 15.039 for 7.5 nm and 10 nm QWs, respectively, using InAs bulk value $\epsilon_{\infty}=14.59$ at the 1.0 eV range (see Ref. \cite{prbdiel}). Although the intrinsic layers include zinc-blende material in the absence of a local field, these values would show infrared regime in the presence of step-by-step changes. It can be concluded that the refractive index yields 3.91 for both sizes, and the decreasing trend occurs when the refractive index takes values in the "resonance" range. Note that the transitions are valid for quantum numbers from $\rm n=1$ to $\rm n=3$, with a rule of $\Delta\rm n=0$.

We can model optical saturation by inserting the spectra mentioned above. Figure \ref{sat4} shows that peak values can be observed through third-order phenomena at the input frequency, $\omega+0\to\omega$. For a transition peak of 0.428 eV and a width of 10 nm, we can observe the familiar trend of decreasing absorption with increasing input intensity at the first heavy-hole exciton. By fitting Eq. (\ref{sat}), we can see optical saturation occurring in the range of $10^3$ $\rm GW/cm^2$. We can then observe that the saturation leads to a value of $I_{\rm s}$=1800 $\rm GW/cm^2$, and the iterative method used in Eq. (\ref{sat3}) has good agreement with familiar saturation fitting.
\section{Conclusion}
This improved treatment shows that the refractive index spectra of the excitonic InAs layers leads to the expected values. In addition to the infrared spectra, optical saturation, which is valid for $\omega+0\to\omega$, can be modeled using this improved approach. Third-order contributions without SHG or THG yield an optical intensity near $10^3$ $\rm GW/cm^2$, demonstrating that the iterative solution provides an effective solution that accounts for nonlinear contributions via third-order phenomena.
\section*{Acknowledgement}
The author wishes to thank the M.Sc. Sefa Ortakaya for helpful library support.


\begin{thebibliography}{99}

\bibitem{foxvb} Mistrik J, Kasap S, Ruda H E, Koughia C and Singh J (2017). Springer Handbook of Electronic and Photonic Materials (Cham: Springer International Publishing) chap Optical Properties of Electronic Materials: Fundamentals and Characterization

\bibitem{chuang}Ahn, D., \& Chuang, S. (1987). Intersubband optical absorption in a quantum well with an applied electric field. Phys. Rev. B, 35, 4149–4151.

\bibitem{fox2} Fox, A. M. (2001). Optical Properties of Solids. Oxford University Press.

\bibitem{s1}Rosencher, E., \& Bois, P. (1991). Model system for optical nonlinearities: Asymmetric quantum wells. Phys. Rev. B, 44, 11315–11327.

\bibitem{s2}Restrepo, R., Castano-Vanegas, L., Martinez-Orozco, J., Morales, A., \& Duque, C. (2019). Mid-Infrared linear optical transitions in $\delta$-doped AlGaAs/GaAs triple-quantum well.. Applied Physics A: Materials Science \& Processing, 125(1).

\bibitem{s4}Amin, N., \& Peter, A. (2022). Structure dependent third order nonlinear susceptibility in the presence of impurity and magnetic field in CdS/ZnS core/shell quantum dot. Physica B: Condensed Matter, 643, 414162.

\bibitem{s5}Tiutiunnyk, A., Duque, C., Caro-Lopera, F., Mora-Ramos, M., \& Correa, J. (2019). Opto-electronic properties of twisted bilayer graphene quantum dots. Physica E Low-Dimensional Systems and Nanostructures, 112, 36–48.

\bibitem{s6}En-nadir, R., El-ghazi, H., Belaid, W., Tihtih, M., Abboudi, H., Maouhoubi, I., Jorio, A., \& Zorkani, I. (2022). Intrasubband-related linear and nonlinear optical absorption in single, double and triple QW: the compositions, temperature and QW’s number effects. Philosophical Magazine, 1–14.

\bibitem{s7}Suman Dahiya, Siddhartha Lahon, \& Rinku Sharma (2023). Physica E: Low-dimensional Systems and Nanostructures, 147, 115620.

\bibitem{a1}Sugawara, M., Fujii, T., Yamazaki, S., \& Nakajima, K. (1990). Theoretical and experimental study of the optical-absorption spectrum of exciton resonance in $\mathrm In_0.53$$\mathrm Ga_0.47$As/InP quantum wells. Phys. Rev. B, 42, 9587–9597.

\bibitem{a2}Zhou, G., \& Runge, P. (2014). Modeling of Multiple-Quantum-Well p-i-n Photodiodes. IEEE Journal of Quantum Electronics, 50(4), 220-227.

\bibitem{a3}Li S. \& Xia J. (1997). Title is required!. Phys. Rev. B, 55, 15434.

\bibitem{a4}Han, X., Li, J., Wu, J., Cong, G., Liu, X., Zhu, Q., \& Wang, Z. (2005). Intersubband optical absorption in quantum dots-in-a-well heterostructures. Journal of Applied Physics, 98(5), 053703.

\bibitem{diel1}Holmström, P., Thylén, L., \& Bratkovsky, A. (2010). Dielectric function of quantum dots in the strong confinement regime. Journal of Applied Physics, 107(6), 064307.

\bibitem{Shen2003}Shen, Y. (2003). The Principles of Nonlinear Optics. Wiley.

\bibitem{prbdiel}Aspnes, D., \& Studna, A. (1983). Dielectric functions and optical parameters of Si, Ge, GaP, GaAs, GaSb, InP, InAs, and InSb from 1.5 to 6.0 eV. Phys. Rev. B, 27, 985–1009.
\end{thebibliography}
\end{document}